\documentclass[]{article}
\renewcommand{\arraystretch}{1.3}

\catcode`\@=11
\def\marginnote#1{}

\newcount\hour
\newcount\minute
\newtoks\amorpm
\hour=\time\divide\hour by60
\minute=\time{\multiply\hour by60 \global\advance\minute by-\hour}
\edef\standardtime{{\ifnum\hour<12 \global\amorpm={am}%
        \else\global\amorpm={pm}\advance\hour by-12 \fi
        \ifnum\hour=0 \hour=12 \fi
        \number\hour:\ifnum\minute<10 0\fi\number\minute\the\amorpm}}
\edef\militarytime{\number\hour:\ifnum\minute<10 0\fi\number\minute}

\def\draftlabel#1{{\@bsphack\if@filesw {\let\thepage\relax
      \xdef\@gtempa{\write\@auxout{\string
          \newlabel{#1}{{\@currentlabel}{\thepage}}}}}\@gtempa \if@nobreak
    \ifvmode\nobreak\fi\fi\fi\@esphack} \gdef\@eqnlabel{#1}} 
    \def\@eqnlabel{}
\def\@vacuum{}
\def\draftmarginnote#1{\marginpar{\raggedright\scriptsize\tt#1}}

\def\draft{
  \oddsidemargin -.5truein
  \def\@oddfoot{\footnotesize \sl preliminary draft \hfil
    \rm\thepage\hfil\sl\today\quad\militarytime}
  \let\@evenfoot\@oddfoot \overfullrule 3pt
    \let\label=\draftlabel
    \let\marginnote=\draftmarginnote
  \def\@eqnnum{(\theequation)\rlap{\kern\marginparsep\tt\@eqnlabel}%
    \global\let\@eqnlabel\@vacuum}

  }

\makeatletter
\newdimen\normalarrayskip              
\newdimen\minarrayskip                 
\normalarrayskip\baselineskip
\minarrayskip\jot
\newif\ifold             \oldtrue            \def\new{\oldfalse}
\def\arraymode{\ifold\relax\else\displaystyle\fi} 
\def\eqnumphantom{\phantom{(\theequation)}}     
\def\@arrayskip{\ifold\baselineskip\z@\lineskip\z@
     \else
     \baselineskip\minarrayskip\lineskip2\minarrayskip\fi}
\def\@arrayclassz{\ifcase \@lastchclass \@acolampacol \or
\@ampacol \or \or \or \@addamp \or
   \@acolampacol \or \@firstampfalse \@acol \fi
\edef\@preamble{\@preamble
  \ifcase \@chnum
     \hfil$\relax\arraymode\@sharp$\hfil
     \or $\relax\arraymode\@sharp$\hfil
     \or \hfil$\relax\arraymode\@sharp$\fi}}
\def\@array[#1]#2{\setbox\@arstrutbox=\hbox{\vrule
     height\arraystretch \ht\strutbox
     depth\arraystretch \dp\strutbox
     width\z@}\@mkpream{#2}\edef\@preamble{\halign
\noexpand\@halignto
\bgroup \tabskip\z@ \@arstrut \@preamble \tabskip\z@ \cr}%
\let\@startpbox\@@startpbox \let\@endpbox\@@endpbox
  \if #1t\vtop \else \if#1b\vbox \else \vcenter \fi\fi
  \bgroup \let\par\relax
  \let\@sharp##\let\protect\relax
  \@arrayskip\@preamble}
\def\eqnarray{\stepcounter{equation}%
              \let\@currentlabel=\theequation
              \global\@eqnswtrue
              \global\@eqcnt\z@
              \tabskip\@centering
              \let\\=\@eqncr
 \halign to \displaywidth\bgroup
    \eqnumphantom\@eqnsel\hskip\@centering
    $\displaystyle \tabskip\z@ {##}$%
    \global\@eqcnt\@ne \hskip 2\arraycolsep
         $\displaystyle\arraymode{##}$\hfil
    \global\@eqcnt\tw@ \hskip 2\arraycolsep
         $\displaystyle\tabskip\z@{##}$\hfil
         \tabskip\@centering
    &{##}\tabskip\z@\cr}
\begingroup\ifx\undefined\newsymbol \else\def\input#1 {\endgroup}\fi
\newfont{\hr}{msbm10}
\newfont{\ams}{msam10}

\voffset= - 0.9in
\hoffset= - 1.0in         

\def\beq{\begin{equation}}
\def\eeq{\end{equation}}
\def\ba{\beq\new\begin{array}{c}}
\def\ea{\end{array}\eeq}
\def\be{\ba}
\def\ee{\ea}

\def\Im{{\rm Im}}
\def\F{{\cal F}}
\def\d{\partial}
\def\N2{${\cal N}=2$}

\begin{document}
\begin{flushright}
ITP-UU-01/20\\
SPIN-01/13\\
FIAN/TD-10/01\\
ITEP/TH-26/01
\end{flushright}
\vspace{0.5cm}

\begin{center}
\renewcommand{\thefootnote}{${\!}^\star$}
{\LARGE \bf Electric-Magnetic Duality and WDVV
Equations\footnote{
Contribution to the proceedings of the workshop {\em Classical 
and Quantum Integrable Systems}, Protvino, 2001 (Theor. Math. Phys., to
appear)} } 
\end{center}
\vspace{0.5cm}

\setcounter{footnote}{0}
\begin{center}
\renewcommand{\thefootnote}{\alph{footnote}}
{\large
B.~de~Wit}\footnote{e-mail:\ bdewit@phys.uu.nl}\\
\medskip
{\em Institute for Theoretical Physics and Spinoza Institute,\\ 
Utrecht University, Utrecht,
    The Netherlands}\\
\bigskip\medskip
{\large A.~Marshakov}\footnote{e-mail:\ mars@lpi.ru, andrei@heron.itep.ru}\\
\medskip
{\em Theory Department, Lebedev Physics Institute, and\\ 
Institute of Theoretical and Experimental Physics, Moscow, Russia}\\
\bigskip\bigskip\medskip
\end{center}

\begin{quotation}
\noindent
We consider the associativity (or WDVV) equations in the form they appear in
Seiberg-Witten theory and prove that they are covariant under generic
electric-magnetic duality transformations. 
We discuss the consequences of this covariance from various perspectives.
\end{quotation}

\section{Introduction}
\setcounter{footnote}{0}

Duality transformations play an important role in modern theoretical
physics. In Seiberg-Witten theory \cite{SW} electric-magnetic duality
(for a recent review of electric-magnetic duality   
see \cite{dWlast} and references therein) is a basic
ingredient in obtaining the exact form of the low-energy effective
action. Hence, duality is a crucial tool in studying 
non-perturbative physics. Any truly non-perturbative 
result should be consistent with electric-magnetic duality.

Based on electric-magnetic duality, Seiberg-Witten theory enables the
determination of the holomorphic function $\F({\bf a})$ in terms of
which the low-energy effective action is encoded. Here $\bf a$  
denotes complex fields associated with the Cartan subalgebra of
the gauge group. The function $\F$ plays the role of a prepotential for the
corresponding special K\"ahler geometry. The construction
involves an auxiliary complex curve, whose moduli space of complex
structures is  identified with the special K\"ahler space\footnote{
  Strictly speaking `special geometry' refers to the K\"ahler
  geometry associated with locally ${\cal N}=2$ supersymmetric Yang-Mills
  theories coupled to Poincar\'e supergravity. In the rigid supersymmetry
  context one sometimes uses the term `rigid special geometry'. Here we
  do not make this distinction.} 
with $\bf a$ playing the role of local coordinates. This
construction can be cast in terms of an integrable system \cite{GKMMM},
identifying $\F({\bf a})$ with (the logarithm of) a tau-function of
the so-called quasiclassical or universal Whitham hierarchy \cite{KriW}, which 
satisfies a set of nontrivial differential equations (see, for 
example, \cite{Mbook} and references therein for the details of this 
correspondence). 

An intriguing example of these equations is the set of
associativity or Witten-Dijkgraaf-Verlinde-Verlinde (WDVV) equations
originally found in the context of $2D$ topological theories
\cite{WDVV}. In \cite{MMM} it was established that the function 
$\F({\bf a})$ associated with the exact solution for pure $SU(N)$
Yang-Mills theory \cite{sun}, satisfies the set of associativity
equations, 
\be
\label{WDVV}
\F_i\cdot \F_j^{-1}\cdot\F_k = \F_k\cdot \F_j^{-1}\cdot\F_i
\qquad \forall i,j,k \;,
\ee
written in terms of the matrices $\|\F_i\|_{jk} \equiv \F_{ijk}$ whose
matrix elements are the third derivatives of $\F({\bf a})$,  
\be
\F_{ijk} = {\d^3\F\over\d a_i\,\d a_j\,\d a_k}\;.
\ee
Later it was shown  that (\ref{WDVV}) holds for all known 
Seiberg-Witten solutions based on hyperelliptic auxiliary curves
\cite{MMMlong} \footnote{
  In this work it was also shown that (\ref{WDVV}) is not
  satisfied by the function $\F({\bf a})$ for softly broken ${\cal N}=4$ 
  Seiberg-Witten theory, a phenomenon that is not yet completely 
  understood.} 
(for recent progress for the case of non-hyperelliptic curves, see
\cite{HKM}). 
A crucial ingredient of this result is that the $\F_{ijk}$ determine
the structure constants of the algebra constructed from holomorphic
differentials on the associated complex curve. Because the effective
low-energy theory is subject to electric-magnetic duality, the WDVV
equations (\ref{WDVV}) should respect this duality. 

We emphasize that one should clearly distinguish between the special
K\"ahler metric, constructed from second derivatives of $\F$, and the
so-called metric of the topological theory \cite{WDVV}, constructed
from the third derivatives of $\F$. In the context of (\ref{WDVV}) it
seems clear that the ``topological metric'' plays no role and neither
does the (non-holomorphic) 
special K\"ahler metric, while the matrix
$\|\F_j\|^{-1}$ can be regarded as the inverse of a (holomorphic) metric
$\F_j$. However, $\|\F_j\|^{-1}$ can equally well be replaced by 
the inverse of any linear combination of the matrices $\F_j$ (we
prefer not to interpret the inverse of any linear combination of the
$\|\F_j\|^{-1}$ as a metric). So the issue of a metric is not
immediately obvious in the context of (\ref{WDVV}). 

In this note we demonstrate that the WDVV equations (\ref{WDVV}) are
covariant under generic electric-magnetic 
duality transformations\footnote{
  The covariance of the WDVV equations under electric-magnetic duality  
  was first noticed in \cite{vando}. Restricted
  duality transformations were considered in \cite{ohta}, but the
  results reported there are not fully in agreement with the results
  of this note.}. 
Specifically, we prove that if a holomorphic
function $\F$ satisfies the WDVV equations, so does its {\em dual}  
function. Technically the proof is simple and it is not restricted to the
Seiberg-Witten case; it is applicable to any solution of WDVV
equations and thus, hopefully, to any tau-function of the Whitham
hierarchy for an allowed class of duality transformations. In
sect.~\ref{ss:wdvvdual} we review the associativity equations and
introduce the dual holomorphic function for the duality that interchanges
electric and magnetic charges. In sect.~\ref{ss:gendual} we prove that
upon generic duality  transformations,  the dual function also
satisfies the WDVV equations. Sect.~\ref{ss:disc} contains some
discussion and outlook.  

\section{WDVV equations for the dual prepotential}
\label{ss:wdvvdual}

In Seiberg-Witten theory the second derivatives of ${\cal F}({\bf a})$
are identified with the period matrix of the corresponding auxiliary
complex curve, 
\be
\label{pemat}
T_{ij} =  {\d^2{\cal F}\over\d a_i\,\d a_j}\,.
\ee
It's imaginary part is equal to the K\"ahler metric, as follows
directly from the K\"ahler potential,
\be
K({\bf a},\bar{\bf a}) = \Im\; \sum_i\,{\bar a}_i\,{\d\F\over\d a_i}\,.
\label{K-pot}
\ee
The derivatives of the period matrix, $\d T_{ij}/\d a_k = \F_{ijk}$,
are the (totally symmetric) holomorphic tensors that appear in the
WDVV equation (\ref{WDVV}). We will 
assume that the matrices $\|\F_i\|$ and/or a linear combination
thereof is nonsingular. In contrast to $2D$ topological field theory 
where one linear combination of the matrices $\|\F_{i}\|$ in
(\ref{WDVV}) can always be chosen constant \cite{WDVV,Dub}, this is
not so for Seiberg-Witten theory.  

Let is now consider the electric-magnetic duality transformation
\be
\label{ad}
a_i\to a^D_i = {\d\F\over\d a_i}\,, \qquad a_i^D \to - a_i = {\d\F^D({\bf
a}^D)\over \d a^D_i}\,,
\ee
with the dual function $\F^D({\bf a}^D)$. As is well-known, this
transformation is effected by a Legendre transform,
\be
\F^D({\bf a}^D) = \F({\bf a}) - \sum_i \, a_i\,a^D_i\,. \label{Legendre}
\ee
Obviously we have 
\be
{\d a^D_i\over \d a_j}= {\d^2\F\over\d a_i\,\d a_j} = T_{ij}\,, \qquad
{\d a_i\over \d a^D_j}=  - {\d^2\F^D\over\d a^D_i\,\d a^D_j}= -
T^D_{ij}\,, 
\ee
so that the dual period matrix $T^D_{ij}$ equals minus the inverse of
the original period matrix (\ref{pemat}), {\it i.e.}, 
\be
\label{ttd}
\sum_j \,T^D_{ij}\, T_{jk} = - \delta_{ik}\,.
\ee

Now consider
\be
\label{fd3}
\|\F^D_i\|_{jk} \equiv \F^D_{ijk}= - {\d T^D_{ij}\over\d a^D_k} = 
{\d^3\F^D\over\d a^D_i\,\d a^D_j\,\d a^D_k}\,.
\ee
It directly follows that\footnote{
  The reader may appreciate the following equation,
  $$
  {\d T^D_{ij}\over\d T_{kl}} =  T^D_{ik}\ T^D_{lj}\,.
  $$ }
\be
{\d T^D_{ij} \over \d a^D_k} = \sum_{l,m,n}\, T^D_{il} \,{\d T_{mn}\over \d
a_l}\,T^D_{nj} \;  {\d a_l \over \d a^D_k} \,. 
\ee
Consequently $\F_{ijk}$ transforms simply as,
\be
{\d^3\F^D\over\d a^D_i\,\d a^D_j\,\d a^D_k}= \sum_{l,m,n} \,
{\d^3\F\over\d a_l\,\d a_m\,\d a_n} \;  {\d a_i \over \d a^D_l} \,{\d
a_j \over \d a^D_m}\, {\d a_k \over \d a^D_n} \;, \label{3df-id}
\ee
or, in matrix form, 
\be
\|\F^D_i\|  = \sum_j \, {\d a_i \over \d a^D_j} \; \|T^D\cdot \F_j\cdot
T^D \|\,. 
\ee
{From} this result it is obvious that the equations (\ref{WDVV}) are
valid for the dual function $\F^D({\bf a}^D)$, because
\be
\F^D_i\cdot(\F^D_j)^{-1}\cdot\F^D_k - (i\leftrightarrow k)  = \sum_{l,m,n}
{\d a_i \over \d a^D_m}\,{\d a_k \over \d a^D_n}\,{\d a^D_j \over \d a_l} 
\,\Big[\F_m\cdot\F_l^{-1}\cdot\F_n - (m\leftrightarrow n) \Big] =0\,,
\label{WDVV-eq}
\ee
where on the right-hand side we made use of (\ref{WDVV}) for all
$l,m,n$. 
\section{Generic duality transformations
\label{ss:gendual}}

The same logic can be applied to generic electric-magnetic
duality transformations forming an arithmetic subgroup of $Sp(2r,{\bf
R})$, which generalize the special duality transformation given in formula
(\ref{ad}). Here $r$ is the rank of the gauge group. At the
perturbative level these transformations are continuous. 
The covariance properties that we are about to establish do not depend
on this feature. 

In the dual basis (denoted by the superscript $S$), we have new
variables ${\bf a}^S$ and a new function $\F^S({\bf a}^S)$, defined
by \cite{speKa,dW1}, 
\be
{\bf a}^S = U\cdot {\bf a} + Z\cdot \left({\d\F\over\d{\bf
a}}\right)\,, \nonumber\\
\left({\d\F^S\over\d{\bf a}^S}\right)=  V\cdot
\left({\d\F\over\d{\bf a}}\right)+ W\cdot {\bf a} \,,
\ee
where the $r\times r$ matrices $U$, $V$, $W$, $Z$ combine into an
$Sp(2r,{\bf Z})$ matrix, by virtue of the relations 
\be
\label{uvwz}
U^t\cdot V-W^t\cdot Z = V\cdot U^t-W\cdot Z^t = 1\,, 
\\
U^t\cdot W = W^t\cdot U\,,\qquad Z^t\cdot V = V^t\cdot Z\,.
\ee
The result analogous to (\ref{Legendre}) reads,
\be
\F^S ({\bf a}^S) =  \F({\bf a})  + {\textstyle {1\over2}} {\bf
a}^t\cdot U^t \cdot W\cdot {\bf a} + {\bf a}^t\cdot W^t \cdot Z \cdot
\left({\d\F\over\d{\bf a}}\right)  + {\textstyle {1\over2}}
\left({\d\F\over\d{\bf a}}\right)^t\cdot Z^t \cdot V\cdot
\left({\d\F\over\d{\bf a}}\right)  \,. \label{fsdu}
\ee
Observe that this represents only a (partial) Legendre transform when
$U^t\cdot W = Z^t\cdot V=0$.

{From} these results one proves that the period matrix, again defined
by (\ref{pemat}), and its dual counterpart, 
\be
{\d^2\F^S\over\d a^S_i\,\d a^S_j} = T^S_{ij}\,,
\ee
are related by 
\be
\label{gendual}
T^S = (V\cdot T + W)\cdot S^{-1}(T)  \,.
\ee
The special K\"ahler metric associated with
the K\"ahler potential (\ref{K-pot}),
\be
G_{\bar \imath j} = {\d^2 K({\bf a},\bar{\bf a})\over \d \bar
a_{\bar\imath} \,\d a_j} = \Im\; T_{ij} \,, \label{Kahme} 
\ee 
transforms as 
\be 
G^S= [S^\dagger]^{-1}(\bar T) \cdot G  \cdot S^{-1}(T)\,.  \label{G-dual}
\ee
Here the matrix $S(T)$ is defined by 
\be
\label{S}
S_{ij}(T) = {\d a_i^S\over d a_j} = \|U + Z\cdot T\|_{ij}\,.
\ee

Now we wish to demonstrate that
the {\em third} derivatives of $\F$ and $\F^S$ remain related just as in
(\ref{3df-id}), {\it i.e.}, 
\be
\label{srel}
\F^S_{ijk} =\sum_{l,m,n} \, \F_{lmn}
\,(S^{-1})_{li}\,(S^{-1})_{mj}\,(S^{-1})_{nk}\,. 
\ee 
or,
\be
{\d^3\F^S\over\d a^S_i\,\d a^S_j\,\d a^S_k}= \sum_{l,m,n}\, 
{\d^3\F\over\d a_l\,\d a_m\,\d a_n} \;  
{\d a_i \over \d a^S_l} \,{\d a_j \over \d a^S_m}\,
{\d a_k \over \d a^S_n} \;, \label{3sf-id}
\ee
This result is known from the literature \cite{dW1} and we will
briefly review it's proof. First one shows that 
\be
\label{dts}
\delta T^S = \left(V - (V\cdot T+W)\cdot S^{-1}\cdot Z\right)\cdot
\delta T\cdot S^{-1} = (S^t)^{-1}\cdot\delta T\cdot S^{-1}\,.
\ee
This result (\ref{dts}) follows directly from the equations
(\ref{gendual}), (\ref{S}) and (\ref{uvwz}). Likewise one shows 
that $S^{-1}(T)\cdot Z$ is a symmetric matrix.

Replacing the variation in (\ref{dts}) by a derivative with
respect to ${\bf a}^S$ and using $\F_{ijk} = \d T_{ij}/ \d a_k$ and
(\ref{S}), one readily proves the 
validity of (\ref{srel}). Along the same line as in
sect.~\ref{ss:wdvvdual}, this then leads to the conclusion that
the WDVV equations (\ref{WDVV}) remain invariant under {\it general}
duality transformations (\ref{gendual}), so that the function  
$\F^S({\bf a}^S)$ satisfies
\be
\label{WDVVS}
\F^S_i\cdot (\F^S_j)^{-1}\cdot\F^S_k = \F^S_k\cdot
(\F^S_j)^{-1}\cdot\F^S_i\,. 
\ee
provided the WDVV equations were valid for the original function
$\F$. Upon setting $U=V=0$ and $Z=-W=1$, the reader can also verify that the
results of the previous section are reproduced. Finally, we note that
most of the transformations of this section concern only the holomorphic
sector of the theory, where the symplectic transformations can be
extended to complex-valued matrices. While this would bring one
outside the strict context of electric-magnetic duality, this
extension may have some relevance in the context of integrable models. 

\section{Discussion
\label{ss:disc}}

According to electric-magnetic duality two dual 
holomorphic functions, $\F$ and $\F^S$, describe the same system and
thus belong to the same equivalence class. As we stressed, this
duality is therefore at the basis of the Seiberg-Witten
theory. Consequently it follows that physically relevant results, when
expressible directly in terms of the function $\F$, 
should hold for all representatives of the equivalence
class. Specifically, $\F$ and $\F^S$ should both satisfy the
corresponding relations. Therefore 
it follows that the associativity equations should hold for all
representatives of a given equivalence class. In this note we have
shown this to be the case as the associativity equations (\ref{WDVV})
are simply covariant under generic duality transformations 

We should point out here that versions of the associativity
equations in Seiberg-Witten theory different from (\ref{WDVV}) have
appeared in the literature (in particular, see
\cite{BonelliMatone}).  The incorrectness of these versions can
be deduced from their lack of covariance with respect to
electric-magnetic duality; hence it should not come as a surprise that
they have meanwhile been rejected on other grounds as well. 

The issue of the metric that appears in the associativity equations
remains a confusing one. We have already stressed that the special
K\"ahler metric has nothing in common with the ``metric'' in the
context of the $2D$ topological theory that underlies the original
WDVV equations. The latter is related to the third derivative of some
function and can be chosen constant. Note that the extra condition of
the constancy is {\em not} preserved under duality, so that duality
seems to take us out of the class of topological solutions in the sense
of \cite{Dub}. The first metric, on the other hand, is related to the
second derivative and it is non-holomorphic and transforms
non-holomorphically under duality (cf. (\ref{G-dual})). In
contradistinction with the above, the
metric in the context of the associativity equations (\ref{WDVV}) is
clearly non-constant and holomorphic. To clarify this issue is
obviously relevant. 

Our result goes beyond Seiberg-Witten theory because it applies to all
cases where the WDVV equations are valid, irrespective of whether one
can identify proper arguments for the relevance of electric-magnetic
duality for the cases at hand. All Seiberg-Witten solutions are related
to integrable systems, 
where the function $\F$ is the logarithm of the tau-function of the
universal Whitham hierarchy (restricted to a finite set of variables). 
However, not all tau-functions correspond to Seiberg-Witten solutions
(as far as we know), and some of those nevertheless satisfy the WDVV
equations (\ref{WDVV}). Hence, by applying duality transformations we
obtain other tau-functions satisfying the WDVV equations, without
having an a priori understanding as to why the duality constitutes an
equivalence relation for these tau-functions. Duality transformations
with $U^t\cdot W = Z^t\cdot V=0$, where (\ref{fsdu}) takes the form of
a (partial) Legendre transform,  may be of particular importance in
the context of the Whitham hierarchies. 

Yet another issue concerns the relation of WDVV equations in
Seiberg-Witten theory with the geometry of moduli spaces of Riemann
surfaces and integrable systems. Certainly dual period matrices are
not distinguishable from the point of view of the geometry of complex
curves. They are equivalent and the corresponding equivalence of the
associativity equations is a consequence of this fact. On the other
hand, it is well-known (see, for example, \cite{Dub,MiModu}) that when
two different functions $F({\bf a})$ and ${\tilde F}(\tilde{\bf
a})$ satisfy 
\be
{\d^2 F\over \d a_i\,\d a_j} = {\d^2 {\tilde F}\over \d {\tilde a}_i\,
\d {\tilde a}_j}\,, \label{DubMiMo}
\ee
and $F({\bf a})$ is a solution to WDVV equations, then ${\tilde
F}(\tilde{\bf a})$ is trivially a solution to the same equations. In
this note we extended this equivalence to the case of functions whose
second derivatives ({\it i.e.} their period matrix) are related by duality
transformations. Observe that, while representing the same geometry
and belonging to the same equivalence class, the two functions which
solve the WDVV equations are in general completely different as
functions depending on their respective arguments.

\section*{Acknowledgements}
We thank Soo-Jong Rey for hospitality extended to us at the Center
for Theoretical Physics of Seoul National University where this work
was completed. A.M. is grateful to Harry Braden for many discussions
about the relation between special K\"ahler geometry and associativity 
equations and to A.~Mironov and A.~Zabrodin for discussing the role of
duality in the context of Whitham hierarchies. This work was
supported in part by INTAS grant No.~99-1-590. A.M. acknowledges
partial support by RFBR grant No.~00-02-16477, CRDF grant No.
RP1-2102 (6531) and the grant for the support of scientific schools
No.~00-15-96566. 

\end{document}